\newcommand{\rem}[1]{}
\documentclass{amsart}
\usepackage{amsfonts,amssymb,amsmath,amsthm,mathrsfs}
\usepackage{url}
\usepackage{graphicx}

\newtheorem{thrm}{Theorem}[section]

\newtheorem{prop}[thrm]{Proposition}

\newtheorem{remark}[thrm]{Remark}
\newtheorem{definition}[thrm]{Definition}

\begin{document}
\author[G. M. Cicuta and L. G. Molinari]
{Giovanni M. Cicuta and Luca Guido Molinari}
\address{Dip. Fisica, Univ. Parma, Parco Area delle Scienze 7A, 43100 Parma, Italy,  --
Physics Department, Universit\`a degli Studi di Milano and  I.N.F.N. sez. Milano, Via Celoria 16, 20133 Milano, Italy}
\email{giovanni.cicuta@fis.unipr.it, luca.molinari@unimi.it}
\title[Random Antagonistic Matrices]{Random Antagonistic matrices}
\begin{abstract}
The ensemble of antagonistic matrices is introduced and studied. In antagonistic matrices 
the entries $\mathcal A_{i,j}$ and $\mathcal A_{j,i}$ are real and have opposite signs, or are both zero, 
and the diagonal is zero. This generalization of antisymmetric matrices is suggested by the linearized dynamics of competitive species in ecology.
\end{abstract}
\maketitle

\section{Introduction}
In the past $60$ years the theory of random matrices had an impressive development 
in theoretical physics and in a variety of disciplines. Further progress and usefulness of 
random matrices will be linked to the ability of a specific class of random matrices to 
encode the relevant properties of a specific problem.

For instance, random matrices with entries that vanish outside a band around the
diagonal have been studied for decades as models for the crossover 
between a strongly disordered insulating regime, with localized eigenfunctions and weak 
eigenvalue correlations, and a weakly disordered metallic regime, with extended 
eigenfunctions and strong eigenvalue repulsion \cite{seli,cas,chir}. Such crossover is believed to occur 
in the spectra of certain random partial differential (or difference) operators as the 
spectral parameter (energy) is changed. A review is \cite{spe}.\\
A very different case, which deserves much study, is the network of neurons. In several 
models the interconnections are represented by a synaptic matrix with elements drawn 
randomly. The distribution of eigenvalues of this matrix is useful in the study of 
spontaneous activity and evoked responses. It was pointed out by Rajan and Abbott 
\cite{raj}  that each node in a synaptic conductivity network is either 
purely excitatory or inhibitory (Dale's Law), which leads to constraints on the signs of 
the matrix elements: all entries in a row describing an excitatory neuron must be positive 
or zero, and all entries in an inhibitory row must be negative or zero. Little is known of 
the generic properties of this ensemble of random matrices \cite{amir}.\\
 
In this paper we study a new class of matrices, here called {\em antagonistic matrices}. 
They are characterized by real entries $\mathcal A_{i,j}$ and $\mathcal A_{j,i}$ 
having opposite signs, for all $i<j$, or both zero, and $\mathcal A_{i,i}=0$. As such, they are a 
generalization of real antisymmetric matrices. An example of order $4$ is
\begin{align*} 
{\mathcal A} = \left (  \begin{array}{cccc} 0 & 5.3 & 0 &-1.7 \\ -3.2 & 0 & 2.3 & 2.0\\  
0 & -8.7 & 0 & -6.3 \\ 1.1 & -1.8 & 1.9 & 0 \end{array} \right)
\end{align*}
The reason for the name and the interest of such matrices is their possible relevance 
in models for competitive species (predator-prey) and for the complexity-stability 
debate or paradox in theoretical ecology \cite{all}, which is here summarized.

In a large island, a large number $n$ of species live. Let $n_i(t)$ be the number of 
living individuals of species $i=1,\dots,n$ at time $t$. Let us  suppose that the 
interactions are described by the model
\begin{align}
 \frac{d\,n_i(t)}{d\,t}=h_i\left( n_1(t), \dots ,n_n(t) \right) \; , \quad  i=1, \dots ,n \label{MAYEQ}
 \end{align}
A stationary feasible configuration, also called equilibrium point, is a configuration 
such that for all species:
\begin{align*}
h_i\left ( n_1^*, \dots ,n_n^* \right)=0 \; , \quad  n^*_i\geq 0.
\end{align*}
Let $x_i (t) =n_i(t)-n_i^*$  represent the deviation from the equilibrium point. 
For small deviations the dynamics is linearized:
\begin{gather}
 \frac{d \, x_i (t) }{d\,t}  
 \sim \sum_{j=1}^n M_{i,j} x_j(t),  \quad 
  M_{i,j} = \frac{ \partial h_i}{\partial n_j}\bigg|_{n_r=n_r^*}
 \end{gather} 
Linear stability of the equilibrium point requires that all the eigenvalues of the matrix $M$ should have negative real part. \\

In general, the matrix $M$ is huge and the entries are almost impossible to quantify.
Robert May \cite{may} considered a model where the diagonal elements are all equal, 
$M_{k, k}=-\mu$, $\mu>0$, and the matrix $\tilde M$ of off-diagonal
elements is a real $n \times n$ random matrix. He chose the entries  
as independent identically distributed (i.i.d.) random variables, the single 
probability density $p(\tilde M_{i,j})$ having zero mean and variance $\sigma^2$. In the 
limit $n \to \infty$, with proper assumptions on the moments 
of the probability law,  the density of eigenvalues of the matrix ${\tilde M}$ converges 
weakly to the uniform distribution on the disk $\{z \in {\mathbb C}, \;
|z| \leq \sigma\sqrt{ n} \}$. This is known as the circular law; a survey is \cite{cha}. \\
Provided that $-\mu+\sigma\sqrt{n}\leq 0$, the eigenvalues of $M$ are predicted with large probability to have negative real part. However, with a fixed 
value $\mu$, a more complex system (that is increasing the number $n$ of interacting species) will have an increasing number of eigenvalues with positive real part, 
and will be linearly unstable. \\
The assertion that the increasing complexity of the ecological system leads to its instability 
was (and is) considered false in view of evidence. The critical analysis of R. May's 
paradox may be found in \cite{ESA,rob,imp,hay,McC,for}.\\

The extreme simplicity of R. May's argument is challenging. Is it possible that all the 
eigenvalues of a matrix $M=D+\tilde M$ with  structure plausible to describe an ecological population, have negative real part?\\
In the mathematical literature, a matrix is said to be stable if its spectrum lies in the open left half-plane (a survey is \cite{hersh}). However, the conditions on the principal minors make this approach
of little use for matrices of large order. The location of eigenvalues in the complex plane 
may be bounded by constraining norms of the matrix or matrix rows or columns \cite{horn}. 
Every norm increases as the size of the matrix increases, suggesting a larger region 
for the location of eigenvalues.\\

An interesting analysis of the dynamics \eqref{MAYEQ}, was done by Fyodorov 
and Khoruzhenko \cite{Fyo2015}. Instead of brute linearisation, they consider
an appropriate non-linear form of the driving forces $h_i(\vec x)$, modelled by
$\vec x$-dependent random matrices. They evaluate the average number
of minima $\vec x^*$ for large $n$ and observe a sharp transition between a 
single minimum and a very large number of minima, stable and unstable, as 
the parameter $\mu $ crosses a critical value. Therefore the dynamics could be not as 
May's catastrophe, but a meandering between minima.\\

In this paper we pursue a route suggested by empirical evidence.
The extensive literature on models of real systems of many species points to three features which increase the stability of the system: 1) the species have a competitive (i.e. antagonistic) interaction: the signs in every pair $M_{i,j}$ 
and $M_{j,i}$ are opposite\footnote{Several species have a mutualistic or cooperative interaction: the signs of the pair $M_{i,j}$ and $M_{j,i}$ are both positive. The stability of large system of mutualistic species seem to be related to a very different structure of the matrix. Mutualistic interactions are ignored in this paper.}; 2) there are weak couplings among several species;
3) the matrix is sparse.
The ensemble of random antagonistic matrices 
may accommodate the three features.\\

Random antagonistic matrices are related to random real antisymmetric matrices and to the elliptic ensembles, whose properties are summarized in Sect.2.
In Sect.3 the new set of antagonistic matrices is introduced, with a discussion of single-matrix properties as well as ensemble properties, with
examples. The last Sect.4 is devoted to "almost antagonist" matrices, where the spectral goal of negative real part of eigenvalues is achieved
and the matrices become strictly antagonistic in the large $n$ limit.

We summarize here the conclusions of this work: ensembles of random antagonistic matrices seem to provide a proper model to describe interactions among antagonistic species 
in ecological systems or possibly in other complex systems. They seem useful for their controlled spectral properties.\\
In this paper some analytic statements show a correspondence between certain functions of antisymmetric matrices and analogous functions of antagonistic matrices.\\

Notation. In this paper $M$ indicates a matrix $n \times n$ with real entries, $D$, $S$, 
$A$ and $\mathcal A$ indicate respectively a real diagonal, symmetric, antisymmetric
and antagonistic matrix (see Sect. 3).

\section{Antisymmetric ensemble, elliptic ensemble, dilute matrices}
\subsection{The antisymmetric ensemble.}
We recall elementary properties of the eigenvalues of a real antisymmetric 
matrix $A$. The non-vanishing eigenvalues are pairs of opposite imaginary numbers. 
If $n$ is odd, zero is always an eigenvalue (with odd multiplicity) and $\det A=0$.
If $n$ is even and if zero is not an eigenvalue, then $\det A>0$. 
We shall find that expectation values of the determinant of random antagonistic matrices 
reproduce these properties.

\begin{figure}[!h]
\centering
\includegraphics[height=3cm]{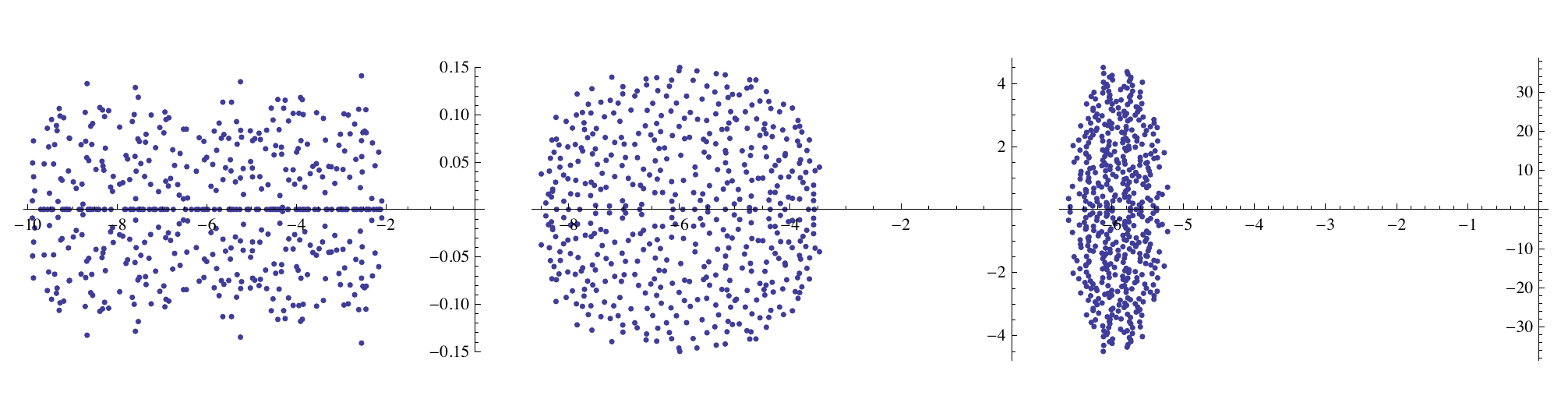}
\caption{\label{F1} Eigenvalues of $M=D+gA$, $n=500$. 
The entries of $D$ are random numbers with uniform probability in $(-10,-2)$, while the 
independent entries of $A$ are random numbers with uniform probability in $(-4,4)$. $g=0.01,\, 0.08,\, 0.5$.}
\end{figure}

\begin{figure}[!h]
\begin{center}
\includegraphics[height=2.5cm]{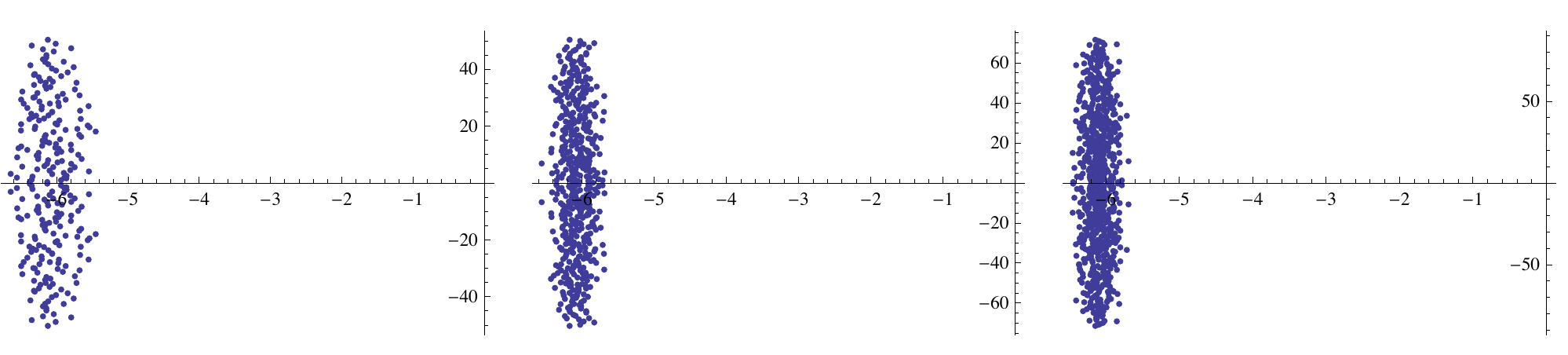} \\
\caption{\label{F2} Eigenvalues of $M=D+gA$, $g=1$, $n=250, \,500, \, 750$.}
\end{center}
\end{figure}

Let us consider matrices $M=D+A$, where $A$ is real antisymmetric and $D$ 
is a real diagonal matrix, with entries in an interval, $a\leq d_j \leq b$. \\
It is known that the eigenvalues $z_k$ of any matrix $Z$ are in the rectangle
Re~$z_k\in \sigma_1$, Im~$z_k\in\sigma_2$ where $\sigma_1$ is the range of $\frac{1}{2}(Z+Z^\dagger)$
and $\sigma_2$ is the range of $\frac{1}{2i}(Z-Z^\dagger)$ (Bendixson, see for ex. \cite{meh}). It follows that
 the eigenvalues $z_k$ of the matrix $M$ are in the rectangular region 
$a\leq Re \, z_k \leq b$ , $-\beta \leq Im\, z_k \leq \beta$ , where $\pm i\beta$ are the extreme eigenvalues of $A$.
Since the result holds for any matrix $M=D+A$, we have:
\begin{prop} \label{prop_2.2} 
Let $D$ be diagonal real random matrices with elements $d_j$ drawn with a probability law such that $a\leq d_j \leq b$ for every $j=1, \dots , n$. Let  $A$ be any real antisymmetric matrix. The eigenvalues of $M=D+A$ are in the strip $a\leq Re \,z \leq b$.
\end{prop}
A simple simulation exhibits the relevant features. The eigenvalues of a random matrix 
$M=D+g\,A$ of order $n=500$, $g=0.01,\, 0.08,\, 0.5$, are depicted in fig.\ref{F1}. 
The panel shows the evolution from the diagonally dominant case to the case where merely the barycenter $x=-6$ of the diagonal matrix affects the dominant  antisymmetric component.
\\
Fig.\ref{F2} again depicts the eigenvalues of $M=D+g\,A$, with the same choice for the entries of $D$ and $A$,  
but with $g$ fixed to $1$ and increasing values of the order $n$.
Only the vertical spread of the eigenvalues increases with $n$.
\\

Clearly, the random matrices $M=D+g\,A$ satisfy the stability requirement that 
all eigenvalues have negative real part, independent of the size of the matrix and of the size 
of the entries of the antisymmetric component. However they would provide a model too rigid to 
describe a realistic community.

\begin{remark} Let's consider again the matrix $M=D+g\,A$, where the 
entries of the diagonal matrix are negative, $a \leq d_j \leq b<0$ for all $j$, 
$A$ is any real antisymmetric matrix. The eigenvalues of $M$ are in the strip $a \leq {\rm Re}\,z_j \leq b$. If $O$ is an orthogonal matrix, consider the new matrix 
$$ M'=O\, (D+g\,A)\, O^{-1}= \bar D +\tilde S+g\,A' $$ 
where $\bar D$ is the diagonal part and $\tilde S$ is the off-diagonal part of the symmetric matrix $O D O^{-1}$.
The entries of $\bar D$ are bounded:
$$a \leq {\bar d}_j=\sum_k O_{j,k}^2 d_k \leq b$$
This suggests a possible structure for a real matrix $M$ with the desired spectral properties: the antisymmetric part   $(M-M^T)/2$ is arbitrary and the symmetric part $ S=(M+M^T)/2={\bar D} +\tilde S$ is diagonally dominant. A simple way to achieve it is to choose the diagonal elements $\bar d_j=S_{j,j}$ in the interval $(a,b)$ and the off-diagonal elements $S_{i,j}=O(1/\sqrt{n})$. This example is made explicit in Sect.4.\\
\end{remark}

\subsection{\bf The Elliptic Ensemble}
The best known elliptic ensemble is a model of random real matrices $J_{i,k}$ with Gaussian probabilities \cite{somm}:
\begin{align}
P(J)\,\prod_{i,j} dJ_{i,j}= \frac{1}{Z}\exp \left[ -\frac{n}{2(1-\tau^2)} 
{\rm Tr} (J\,J^T-\tau \,J\,J)\right ] \,\prod_{i,j} dJ_{i,j} , \quad |\tau|\le 1 
\end{align}
By writing $J=S+A$ with  $S=\frac{1}{2} (J+J^T)$ and  $A=\frac{1}{2}(J-J^T)$ one evaluates
${\rm Tr} (J\,J^T) = {\rm Tr}(S^2-A^2)$ and ${\rm Tr}(J\,J) ={\rm Tr}(S^2+A^2)$. Then
\begin{gather*}
P(S+A) = \frac{1}{Z} \left[ \prod_{i=1}^n e^{-\frac{n}{2(1+\tau)} (S_{i,i})^2}\right ]
 \left[  \prod_{i=1}^n \prod_{k>i} e^{-\frac{n}{1+\tau}(S_{i,k})^2}\right] 
		\left[ \prod_{i=1}^n \prod_{k>i}  e^{-\frac{n}{1-\tau}(A_{i,k})^2}\right ]\\
Z= \left( \frac{\pi}{n}\right)^{n^2} \left[2(1+\tau)\right]^n\left[1-\tau^2\right]^{n(n-1)/2}
\end{gather*}
The set of $n^2$ random real variables is partitioned into three sets of independent 
central normal random variables: $n$ variables $S_{i,i}$ with $\sigma^2=(1+\tau)/n$,
$\frac{n(n-1)}{2}$ variables $S_{i,k}$  ($i<k$) with $\sigma^2=(1+\tau)/(2n)$, and
$\frac{n(n-1)}{2}$ variables $A_{i,k}$ ($i<k$) with $\sigma^2=(1-\tau)/(2n)$. One 
evaluates
$$\mathbb E\left [ J_{i,k}\right] = 0, \quad \mathbb E\left[J_{i,k} J_{k,i}\right]=  \frac{\tau}{n},
\quad \mathbb E\left[(J_{i,k})^2\right] = \frac{1}{n}. $$
In the limit $n\to\infty$, the distribution of the eigenvalues of $J_n \sqrt n$ 
converges to the uniform distribution on the elliptic region with semi-axes 
$a=(1+\tau)\sqrt{n}$, $b=(1-\tau)\sqrt{n}$.\\

Some decades of progress are evident in the more recent works \cite{nau,ngu}. The
following theorem is a generalization of the Circular Theorem, by Girko and Ginibre,
and is important for the present discussion.
\begin{thrm}[Elliptic theorem]\label{prop_2.1}
Let $M$ be a real random matrix such that: \\
a) pairs $\{M_{i,j} \, , \,M_{j,i}\}$, ${i\neq j}$, are i.i.d. random vectors and
$$ \mathbb E (M_{1,2}\, M_{2,1})=\rho,\quad  |\rho|\leq 1$$
b) $\mathbb E (M_{1,2})=\mathbb E(M_{2,1})=0$, $\mathbb E (M_{1,2}^2)=\mathbb E(M_{2,1}^2)=1$,  $\mathbb E (M_{1,2}^4), \mathbb E (M_{2,1}^4)\leq C$;\\
c) The diagonal entries $m_{i,i}$ are i.i.d. random variables with 
$$\mathbb E (M_{1,1})=0\; , \quad \mathbb E (M_{1,1}^2)<\infty $$
Then the distribution of the eigenvalues $x_k+iy_k$ of the matrix $\frac{1}{\sqrt{ n}}M$ converges, in the limit $n \to \infty$, to the uniform distribution on the ellipse
$$  \frac{x^2}{(1+\rho)^2}+\frac{y^2}{(1-\rho)^2} \leq 1 $$
\end{thrm}


\subsection{\bf Dilute matrices}
In realistic models the different species do not have all-to-all connectivity. We should expect
most of the matrix elements of $M$ to vanish. As one introduces an increasing number of zero 
entries, the circular law continues to hold, up to a point.\\
Let us suppose that the $n^2$ real entries $M_{i,j}$ of the matrix $M$ are i.i.d. random variables with a probability $1-Q_n$ to be zero: 
$$P\left(M_{i,j}\right)=Q_n\, \pi(M_{i,j}) + (1-Q_n)\,\delta (M_{i,j})  $$
where $\pi(M_{i,j})$ is a probability distribution with variance $\sigma^2$. \\
If $0 < Q_n<1-\frac{1}{n^{1-\alpha}}$, $0<\alpha \leq 1$, the eigenvalues of 
$M$ converge to the uniform distribution on a disk of radius $\sigma\,\sqrt{n\,Q_n}$ \cite{wood}.\\
If $Q_n=p/n$, the graph associated to the matrix typically decomposes into a giant cluster ($p=1$ is a percolation transition) and a large number of small clusters, mostly trees. 
The spectral density of eigenvalues shows spikes corresponding to the eigenvalues of trees \cite{rod,fyo,bauer,seme,kuhn,tim}.

Neri and Metz \cite{Metz} obtained the analytic form of the spectrum
of diluted random matrices with entries $\{M_{i,j}, M_{j,i}\}$ that take values
$(a,b)$, $(b,a)$ or $(0,0)$ according to a tuneable
hierarchic structure of the graph of the matrix. In the dense limit it recovers
the elliptic distribution.

S. Allesina and  Si Tang \cite{all} correctly 
argued that to consider elliptic ensembles (with the antisymmetric part greater than the symmetric part) with an amount of dilution increases the stability of the system. 
Still the axes of the ellipse are proportional to $\sqrt{n}$ and, for sufficiently large $n$ and if the center of the ellipse is kept fixed at a real negative value, the elliptic domain will not be confined to the left complex half-plane.

\section{\bf Antagonistic matrices}
The goal of this investigation is to explore a class of real matrices useful to describe an ecological community, such that the real part of all the eigenvalues of the matrix is negative 
despite $n$ being large, in order to evade the stability-complexity paradox.
\begin{definition}
An antagonistic matrix ${\mathcal A}$ is a real $n\times n$ matrix such that $\mathcal A_{i,i}=0$
and, for every pair $i<j$, the entries $\mathcal A_{i,j}$ and ${\mathcal A}_{j,i}$ 
have opposite sign or are both zero. 
\end{definition}
\begin{remark}
If $\mathcal A$ is an antagonistic matrix, also $-\mathcal A$ and $\mathcal A^T$ are antagonistic. If $D$ is real diagonal then $D\mathcal A D^{-1}$ is antagonistic. 
If $P$ is a permutation matrix\footnote{In a permutation matrix there is exactly one entry equal to 1 in each row and in each column equal, all other entries are zero}, also $P^T{\mathcal A} P$ is antagonistic, 
with same eigenvalues.
\end{remark}
In ref.\cite{bloch}, it was shown by standard perturbation methods that the non-degenerate 
spectrum of a real symmetric matrix perturbed by an antisymmetric matrix is squeezed to
a narrower rectangle in the complex plane. We show a similar result: 

\begin{prop}\label{3.1} 
Let $D$ be a real diagonal matrix with entries $d_1, \dots, d_n$, with non degenerate 
extremal values $d_M$ and $d_m$, and let ${\mathcal A}$ be an antagonistic matrix. For 
small $\epsilon$ and at leading order, the eigenvalues of $M(\epsilon)= D+\epsilon \, {\mathcal A}$ are in the strip 
\begin{align}
 d_m+ \epsilon^2 \frac{|\mathcal A^2_{m,m}|}{d_M-d_m}
  < z <d_M - \, \epsilon^2 \frac{|\mathcal A^2_{M,M}|}{d_M-d_m}
 \end{align}
\begin{proof} Let us expand in $\epsilon $ the characteristic polynomial:
\begin{align}
P(z,\epsilon ) &= \det (z-D-\epsilon \mathcal A) \nonumber\\
&= P(z,0) \exp [{\rm tr}\log ( 1 - \epsilon  (z-D)^{-1}\mathcal A )]\nonumber \\
&=P(z,0) \left[ 1-\frac{\epsilon^2}{2}\sum_{i,j} \frac{\mathcal A_{i,j}\mathcal A_{j,i}}{(z-d_i)(z-d_j)}
 + \mathcal O (\epsilon^3) \right ] \label{Pz}
\end{align}
The term linear in $\epsilon $ is zero because $\mathcal A_{j,j}=0$. To leading order, 
the extremal eigenvalues of the perturbed matrix $M(\epsilon)$ are:
\begin{align}
\lambda_{\max} = d_M - \epsilon^2 \sum_{j\neq M} \frac{|\mathcal A_{M,j}\mathcal A_{j,M}|}{d_M-d_j} , \qquad
 \lambda_{\min} = d_m + \epsilon^2 \sum_{j\neq m} \frac{|\mathcal A_{m,j}\mathcal A_{j,m}|}{d_j-d_m}
\end{align}
The result follows by a simple inequality. 
\end{proof}
\end{prop}


\begin{prop} \label{3.2} 
With the same setting of proposition \ref{3.1}, let the lowest eigenvalue 
$d_{\min}$ of $D$ have degeneracy $h$. Then a pair of eigenvalues of $M(\epsilon)$
are complex conjugate, and $h-2$ are unperturbed at order $\epsilon$.
\begin{proof}
Let $\sigma $ be the set of $h$ indices such that $d_j = d_{min}$.
The expansion \eqref{Pz} is
\begin{align*}
P(z,\epsilon) =&P(z,0) \left[ 1-\frac{\epsilon^2}{2} \sum_{i,j\in \sigma}\frac{\mathcal A_{i,j}
\mathcal A_{j,i}}{(z-d_{\min})^2} -\epsilon^2 \sum_{i\in\sigma}\sum_{j\notin\sigma} 
\frac{\mathcal A_{i,j}\mathcal A_{j,i}}{(z-d_{\min})(z-d_j)} +\dots \right]
\end{align*}
The solution of $P(z,\epsilon)=0$ for $z=d_{\min}+\epsilon \delta_1+\epsilon^2\delta_2 +\dots $
shows that $h-2$ minimal eigenvalues remain unchanged and two become complex:
 $$\lambda = 
 d_{\min}\pm i \frac{|\epsilon|}{\sqrt 2} \sqrt{\sum_{i,j\in\sigma} |\mathcal A_{i,j}\mathcal A_{j,i}|}+
 \frac{\epsilon^2}{2} \sum_{i\in\sigma}\sum_{j\notin\sigma} \frac{\mathcal A_{i,j}\mathcal A_{j,i}}{d_{\min}-d_j} +
 \mathcal O(\epsilon^3) $$
 Note that $\delta_2>0$. A similar result would hold for a degenerate highest eigenvalue. 
\end{proof}
\end{prop}

\subsection{Random antagonistic ensembles}
The simplest model of an ensemble of random antagonistic matrices has a joint probability density for the $n^2-n$ matrix entries ${\mathcal A}_{i,j}$ in the form of a product of joint probability 
densities for the pairs, i.e. the pairs are independent random vectors, like in the elliptic ensemble:
\begin{align}
 P\left({\mathcal A}\right)= \prod_{i< j} f_{i,j} ({\mathcal A}_{i,j} \, , \, {\mathcal A}_{j,i}) 
 \end{align}
 where $ f_{i,j} (x,y) = P(\mathcal A_{i,j}=x, \mathcal A_{j,i}=y) $. If $f_{i,j}(x,y)= f_{i,j}(y,x)$, the resulting marginal probabilities $p(\mathcal A_{i,j}=x)$ and $p(A_{j,i}=y)$ are  equal.\\
The support of each pair density $f_{i,j}$ is a subset in $(x,y)$ plane where $x\, y\le 0$.
This constraint increases the stability of the model because it increases the weight of the antisymmetric component versus the symmetric component.\\
If $\pm \mathcal A$ belong to the ensemble with the same probability, it follows that if $z$ 
belongs to the spectrum of the ensemble, then the four points $\pm z$ and $\pm z^*$ 
belong to it with same probability.

\begin{remark}
If the independent random pairs are chosen to be identically distributed and the random 
antagonistic model satisfies the conditions of theorem \ref{prop_2.1} then, in the large $n$ limit, 
the eigenvalues converge to a (slim) ellipse.\\ 
For the purpose of stability it is necessary to choose different probability distribution for the pairs.
\end{remark}

The following proposition is reminiscent of the known property of a real antisymmetric matrix $A$: 
$\det A=0$ ($n$ odd), $\det A=(pf [A])^2=-pf [A] \cdot pf [A^T]$ ($n$ even).\\
We recall the notion of Pfaffian. Let $n$ be even. Given a triangular array  $a=\{a_{i,j} \}$, $1 \leq i<j\leq n$,
$$pf[a]= \sum_P{}^\prime \epsilon_P \, a_{i_1,i_2} a_{i_3,i_4}\dots a_{i_{n-1}, i_n}  $$
where the sum is on all permutations $P=
\left( \begin{array}{cccc} 1 & 2 & \dots & n \\ i_1 & i_2 & \dots & i_n \end{array}\right)$ 
such that\\
$$ i_1<i_2, \, i_3 < i_4, \,\dots , i_{n-1}<i_n, \quad \text{and} \quad i_1<i_3<i_5<\dots < i_{n-1}$$ 
$\epsilon_P$ is the sign of the permutation. If $n$ is odd, $pf[a]=0$ by definition.\\
In the case of a square matrix $M$, $pf[M]$ is a multinomial in the entries of the triangular array $\{M_{i,j}\}$, $1 \leq i<j\leq n$. If $A$ is a real antisymmetric matrix, $pf[A]=-pf[A^T]$.
\begin{prop} \label{3.3} 
Let ${\mathcal A}$ belong to a random antagonistic ensemble where the joint probability density of the entries is the product of probability of independent pairs, as in eq.(5)
and the average of each entry is zero, $\mathbb E \left[  \mathcal A_{i,j} \right ]=0$. Then: 
\begin{align*}
&\mathbb E \left[ \det  \mathcal A \right ]=0   &\text{$n$ \rm {odd}} \\
&\mathbb E \left[\det \mathcal A \right ]=(-1)^{n/2} \mathbb E \left[\,pf[{\mathcal A}]\,\cdot \,pf[{\mathcal A}^T]\,\right]>0  &\text{$n$ \rm{even}}
\end{align*}
The expectation of the characteristic polynomial $\mathbb E (\det[z\,I_n-\mathcal A]) $ 
is a polynomial in $z^2$ with positive coefficients. In particular, $
\mathbb E(\sum_k \lambda^2_k) =\mathbb E({\rm tr}[\mathcal A^2]) = -\sum_{i<j} \theta_{i,j}$
where $\theta_{i,j} = - \mathbb E(\mathcal A_{i,j}\mathcal A_{j,i})\ge 0.$
\end{prop}
\noindent
The proofs with the explicit expressions of the average Pfaffian or characteristic polynomial are given in the appendix, with two different techniques.

\subsection{Simple probability measures and spectral domains}
We briefly describe some simple probability densities for random antagonistic matrices, yielding the most common marginal probabilities. 
In the first three examples the independent pairs are identically distributed, $f_{i,j}(x,y)=f(x,y)$.
\subsubsection{\bf Gaussian marginal probability}
\begin{align*}
f (x,y) = \frac{1}{\pi} e^{-\frac{1}{2}(x^2+y^2)} \theta (-xy) 
\end{align*}
The marginal probabilities are standard normal 
$$f(x)=\frac{1}{\pi}\int_{-\infty}^\infty dy\,e^{-(x^2+y^2)/2}\theta(-x y)=\frac{1}{\sqrt{2\pi}}e^{-x^2/2}$$
and $\mathbb E[ {\mathcal A}_{i,j}]=0$, $\mathbb E[( {\mathcal A}_{i,j})^2]=1$,  
$\mathbb E[ {\mathcal A}_{i,j} {\mathcal A}_{j,i}]=-\frac{2}{\pi}$.
\subsubsection{\bf Uniform marginal probability }
\begin{align*}
f(x,y)=  \begin{cases} 1/2 & \text{if $x \in (0,1)$ and $y \in (-1,0)$} \\
1/2 & \text{if $x\in (-1,0)$ and $y\in (-1,0)$}\\
0 & \text{otherwise} \end{cases}
\end{align*}
The marginal probabilities are uniform in $(-1,1) $:
\begin{align*}
f(x)=\int_{-1}^1  f(x , y)\,dy = \begin{cases} 1/2 & \text{if  $x\in (-1,1)$} \\
0  & \text{otherwise}
\end{cases} 
\end{align*}
Each random variable ${\mathcal A}_{i,k} $ is identically distributed, with 
$\mathbb E[\mathcal A_{i,j}]=0$, $\mathbb E[(\mathcal A_{i,j})^2]=\frac{1}{3}$, 
$\mathbb E[(\mathcal A_{i,j})^4]=\frac{1}{5}$ and, for every pair, $\mathbb E [{\mathcal A}_{i,k}  {\mathcal A}_{k,i} ]= -\frac{1}{4}$.

\subsubsection{\bf Marginal probability with support on two symmetric intervals}
If the joint probability density for a pair has support on strips, the 
marginal probability has support on two disjoint intervals.\\
For example, let us define the function 
$$g_w(x)=\begin{cases} \frac{1}{2w} & \text{if $-w<x<w$}\\ 
0 &  \text{ otherwise} \end{cases}  \quad 0<w<1 $$
and the joint probability density of the pair  
\begin{align*}
f( x,y)=\tfrac{1}{2} \left[ g_w (x+1)\, g_w (y-1) +
 g_w(x -1) g_w(y+1) \right] 
\end{align*}
and $\mathbb E\left[ {\mathcal A}_{i,k} {\mathcal A}_{k,i}\right]= -1$.
The marginal densities are 
$ f(x)=\tfrac{1}{2}\left[  g_w(x-1)+  g_w(x+1) \right] $,
that is, the probability density of any ${\mathcal A}_{i,k} $ has support on the union of two 
intervals, $(-1-w,-1+w)\,\cup\, (1-w,1+w)$, with
$\mathbb E[\mathcal A_{i,j}]=0$, $\mathbb E[(\mathcal A_{i,j})^2]=1+\frac{w^2}{3}$.

\begin{remark} The three models agree with the conditions in Proposition \ref{prop_2.1}. 
The parameter $\rho $ describing the elliptic domain of the spectrum in the limit $n\to\infty$ of the random antagonistic matrix is
\begin{align*}
\rho =  \begin{cases}  -\frac{2}{\pi} & \text{\rm Gaussian} \\
 -\frac{3}{4} & \text{\rm uniform} \\
-\frac{3}{3+w^2} & \text{\rm two intervals} \end{cases}
\end{align*}
\end{remark}

\subsection{\bf Independent pairs not-identically distributed.}
\begin{center}
\begin{figure}[!h]
\includegraphics[height=5cm]{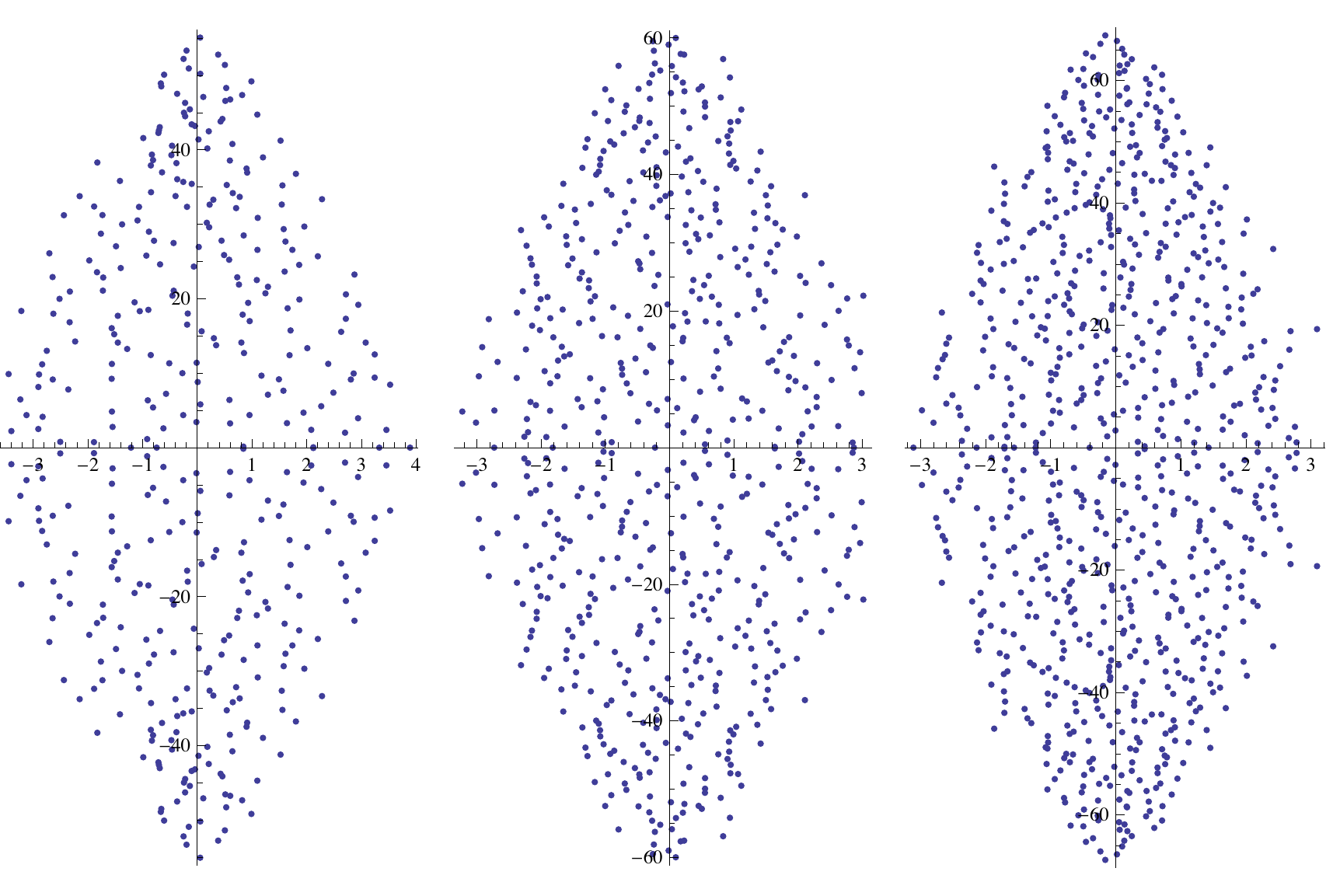} 
\caption{\label{XXX} Eigenvalues of a random antagonistic matrix with $c=50$, $p=8$, $n=400,\, 600, \, 800.$}
\end{figure}
\end{center}

\begin{center}
\begin{figure}[!h]
\includegraphics[height=5cm]{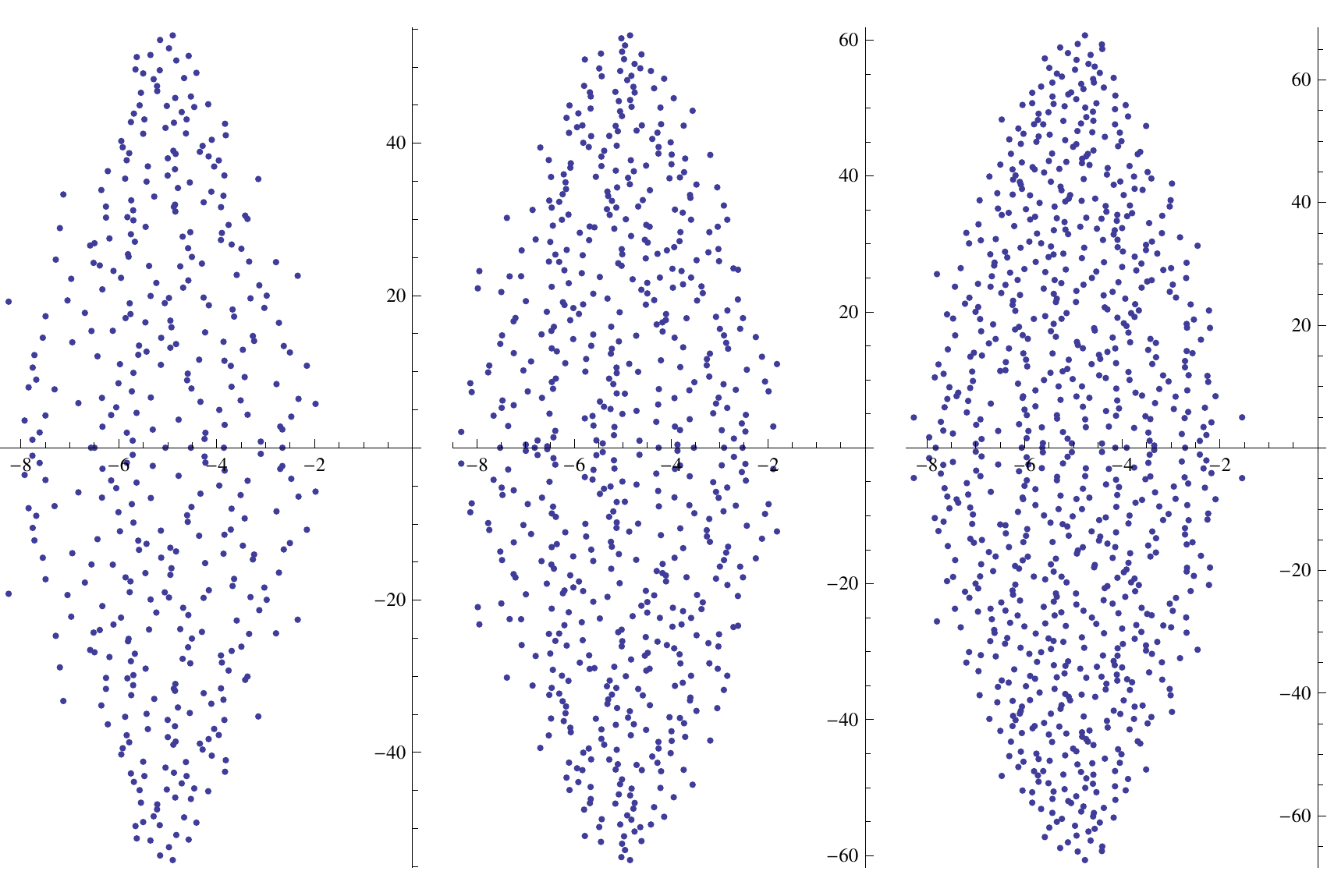}
\caption{\label{XCX} The eigenvalues of $D+\mathcal A$,  $n=400,\,600,\,800$.}
\end{figure}
\end{center}

A useful probability density for the antagonistic matrix  is
\begin{align*}
f_{i,k}(x,y)= \begin{cases}  C_{i,k} & \text{if $x\in (1, 1+\delta )$ and $y\in (-1-\delta, -1)$} \\ 
C_{i,k} & \text{if $y\in (1, 1+\delta )$ and $x\in (-1-\delta, -1)$ } \\ 
0 & \text {otherwise}
\end{cases}, \quad \delta = \frac{c}{1+(k-i)^p}
\end{align*}
\newpage
As the order $n$ of the antagonistic matrix ${\mathcal A}$ increases, the pair of entries far from the diagonal are increasingly similar to an antisymmetric matrix, see fig.\ref{XXX}.
All eigenvalues are in a strip $-a < Re\, z <a$ where the width of the strip does not increase 
with $n$; actually it slightly decreases.
Next, we add a diagonal matrix, with random entries $d_j$ uniform in $(-6,-4)$. 
The eigenvalues of $D+{\mathcal A}$ are shown in fig.\ref{XCX}. They are like the plots in fig.\ref{XXX}, but
shifted by 5 units to the left in the complex plane.
\section{\bf ``Small'' symmetric plus ``big'' antisymmetric.}
A simple way to define an ensemble of real random matrices with eigenvalues that {\em with high probability} are in the 
le ft complex half-plane, is to consider real matrices 
$$D+\frac{1}{\sqrt n} S+A  $$ 
where the symmetric matrix $S$ has zero diagonal and is ``small" compared to the antisymmetric matrix $A$, and $D$ is a properly chosen diagonal matrix. 

\begin{center}
\begin{figure}[!h]
\includegraphics[height=4.5cm]{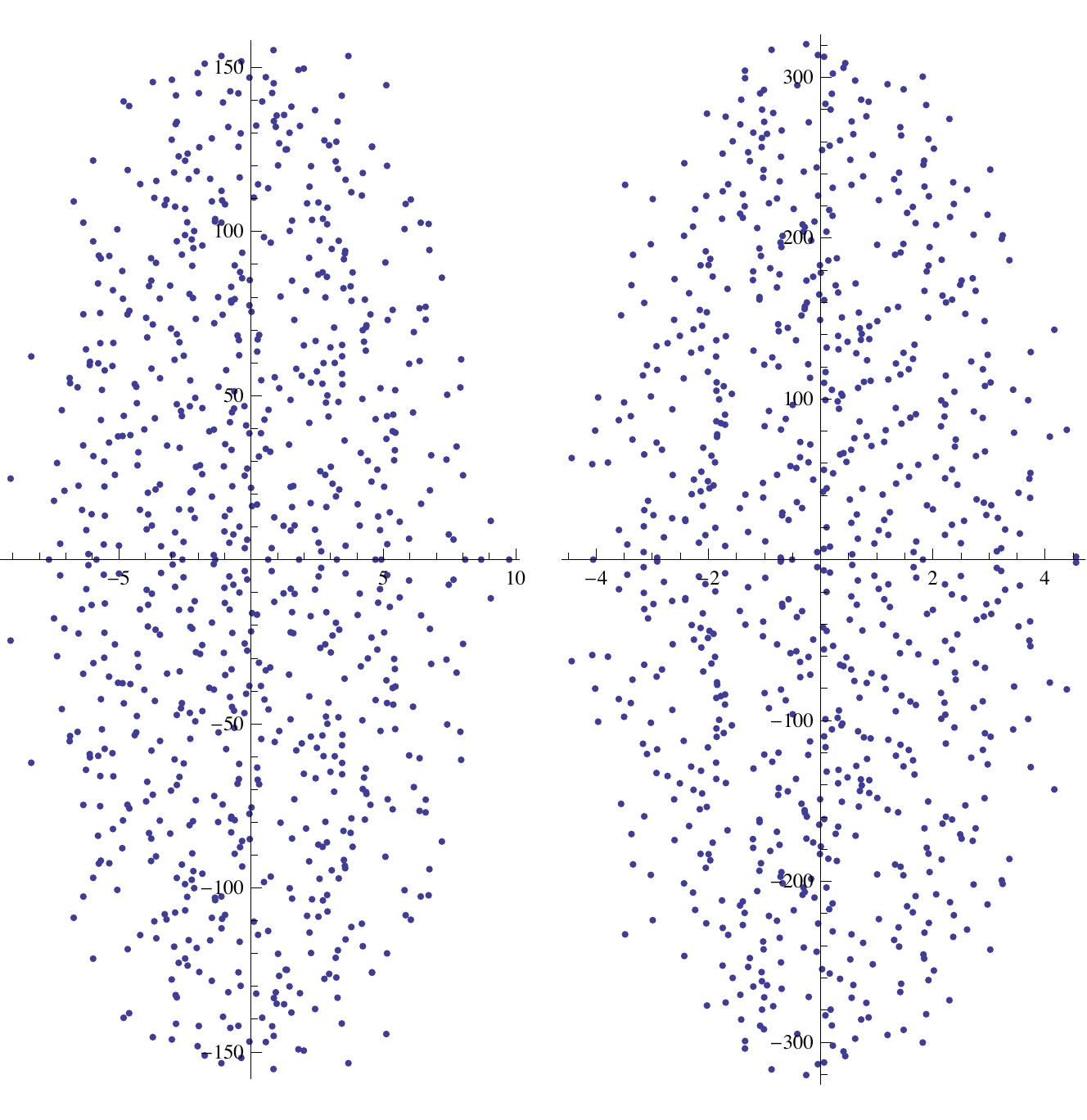}
\caption{\label{DDD} Left: combined eigenvalues of $4$ random matrices $n=200$. Right: eigenvalues of a single random matrix $n=800$. }
\end{figure}
\end{center}

If the entries $S_{i,j}$, for $i>j$, are i.i.d. with zero mean and variance $\sigma_S^2$ , most of the eigenvalues of the matrix  $S/\sqrt{n}$ are, for large $n$, in the interval $(-2\sigma_S\, ,\, 2 \sigma_S)$.\\
If the entries $A_{i,j}$, for $i>j$, are i.i.d. with zero mean and variance $\sigma_A^2$, most of its eigenvalues, for large $n$, are in the interval $(-2i\sigma_A \sqrt{n}\, ,\, 2i \sigma_A \sqrt{n})$.\\
Therefore, for large $n$, the eigenvalues of the matrix $\frac{1}{\sqrt n} S +A$ are with high probability inside the rectangular box with fixed horizontal side $-2\sigma_S\, <x<\, 2 \sigma_S$ and increasing vertical side $-2i\sigma_A \sqrt{n}\,<y< \, 2i \sigma_A \sqrt{n}$.\\ 
Fig.\ref{DDD} shows $800$ eigenvalues of a matrix $S/\sqrt{n}+A$  with entries $S_{i,j}$ ($i>j$)  uniformly distributed in $(-30,30)$ (then $\sigma_s=30/\sqrt{3}$),
and  $A_{i,j}$ ($i>j$) uniformly distributed in $(-10,10)$ (then $\sigma_A=10/\sqrt{3}$).\\
Already for $n=200$ the eigenvalues appear to be confined in a rectangular domain with sides $20 \times 160$, much smaller then the estimated rectangular domain with sides $\left( \frac{120}{\sqrt{3} }\sim 69.2 \right) \, \times \left(\frac{40 \sqrt{200}}{\sqrt{3} }\sim 326.6\right)$.\\
Furthermore the right panel in fig.\ref{DDD} shows that the horizontal side of the domain \textbf{decreases} for increasing values of $n$. This shrinking effect is analogous to that shown in Section 2.1 for the matrix $D+g\,A$.\\
One may also remark that with the above distribution for $S_{i,j}$ and $A_{i,j}$ the random matrix  $\frac{1}{\sqrt n}S+A$ is not antagonistic, but it is antagonistic if the distribution of the $A_{i,j}$ is chosen to have a gap, for instance uniform distribution on $(-10,-1.5) \cup (1.5,10)$ , then $\max |S_{i,j}|/\sqrt{n} <1.5$ for $n\geq 800$ .

\begin{center}
\begin{figure}[h]
\includegraphics[height=4.5cm]{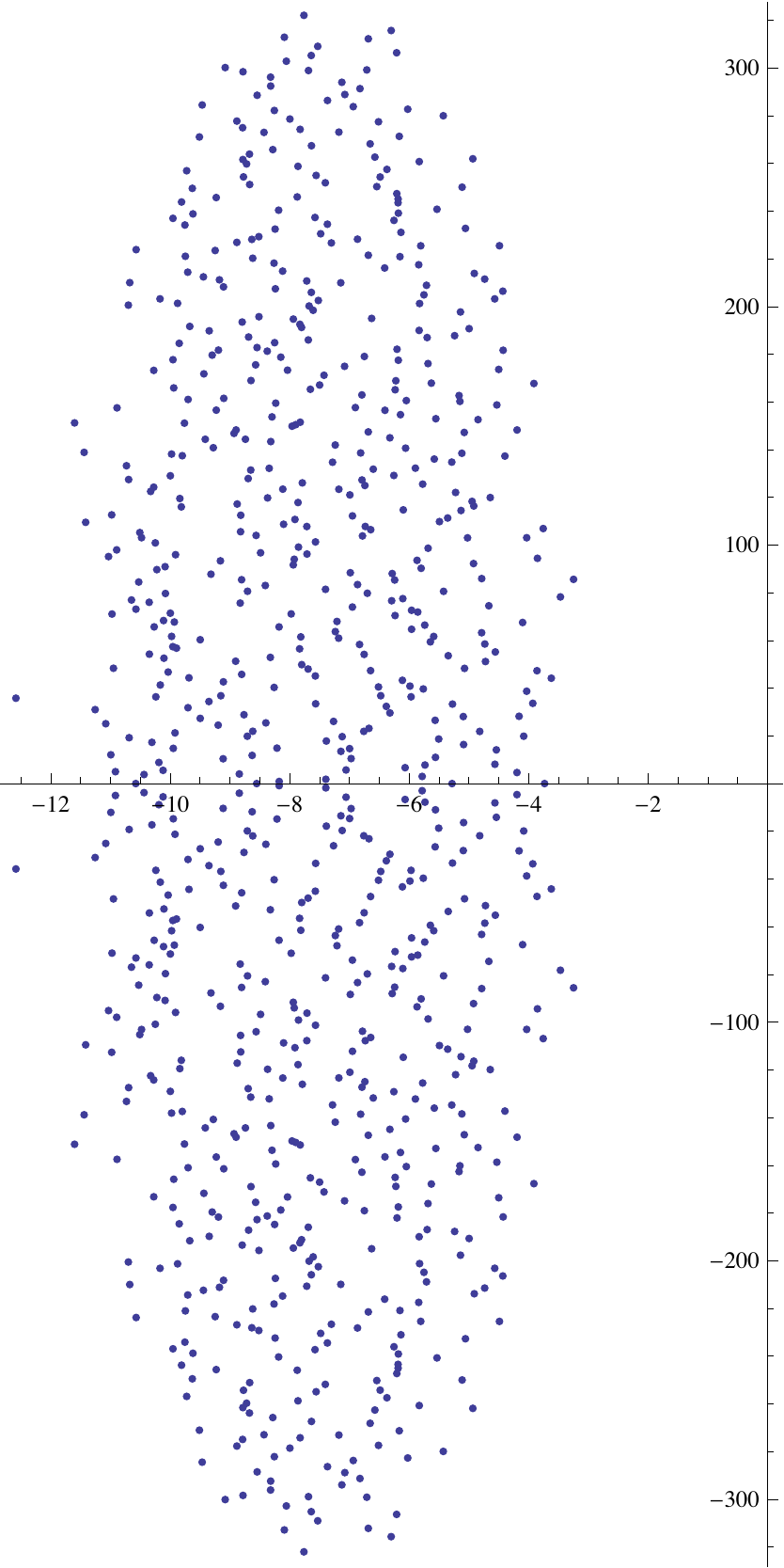} 
\caption{\label{EEE} }
\end{figure}
\end{center}

Finally we add a diagonal matrix: $D+\frac{1}{\sqrt n}S +A$ . With proper choice of  $D$, the domain is shifted so that all eigenvalues are, \textsl{with high probability} in the left part of the complex plane. Fig.\ref{EEE} shows the eigenvalues for $n=800$, with $\frac{1}{\sqrt n}S+A$ having the above distribution and the diagonal entries $d_j$ being uniformly distributed in $(-10,-5)$. All the eigenvalues of the simulation have the real part in the interval $-12.58<x<-3.246$.  It  is reasonable to expect 
that for greater values of $n$ the eigenvalues would be confined into a more narrow strip centered around $x \sim -7.5$.
\section{Appendix}
{\bf Proof 1 (combinatorial).} 
$\mathbb E \left[\det {\mathcal A} \right]=\sum_P \epsilon_P \mathbb E \left [\, \mathcal A_{1,i_1}\dots \mathcal A_{n,i_n}\right]$. 
In analogy with Wick's theorem, the expectation of each term of the sum factorizes and is non-zero only if $n$ is even, and if for every factor $\mathcal A_{k,n_k}$ 
there is the symmetric factor $\mathcal A_{n_k,k}$. 
For example, for $n=4$ the non-zero terms are:
\begin{align*}
\mathbb E \left[\det \mathcal A\right]= \mathbb E\left[\mathcal A_{1,2}\mathcal A_{2,1}\mathcal A_{3,4}\mathcal A_{4,3}+\mathcal A_{1,3}\mathcal A_{2,4}\mathcal A_{3,1}\mathcal A_{4,2}+\mathcal A_{1,4}\mathcal A_{2,3}\mathcal A_{3,2}\mathcal A_{4,1}\right]>0.
\end{align*}
The expectation is non-vanishing only for the permutations which are products of $n/2$ cycles of length two. The number $c_n$ of terms that contribute to
the expectation value of  $\det[{\mathcal A}]$ is\footnote{See for example R.~P.~Stanley, \textsl{Enumerative Combinatorics}, vol.1, pag.18, Cambridge Univ. Press}
\begin{align*}
c_n=(n-1)(n-3)\dots 3\cdot 1= \frac{(n)!}{2^{n/2}\left(\frac{n}{2}\right)!}
\end{align*}
The sign of such permutations is $\epsilon_P=(-1)^{3n/2}$ {}\footnote{See for example: M.~Mahajan, V.~Vinay, \textsl{Determinant: Old Algorithms, New Insights}, 
Electronic Colloquium on Computational Complexity, Report 12 (1998) or 
G.~Rote, \textsl{Division-Free Algorithms for the determinant and the pfaffian: algebraic and combinatorial approaches}, 
Computational Discrete Mathematics 2001.}.\\	
For an antagonistic matrix, the Pfaffian is multilinear in the entries of the upper triangular part of the matrix, $\mathcal A_{i,j}$ with $i<j$. For example ($n=4$):
$$ pf\left[ {\mathcal A}\right] =  \mathcal A_{1,2}\mathcal A_{3,4}-\mathcal A_{1,3}\mathcal A_{2,4}+\mathcal A_{1,4}\mathcal A_{2,3}, \quad pf\left[ {\mathcal A}^T\right] =  \mathcal A_{2,1}\mathcal A_{4,3}-\mathcal A_{3,1}\mathcal A_{4,2}+\mathcal A_{4,1}\mathcal A_{3,2} $$
If $n$ is even, the number of terms in $pf \left [ \mathcal A\right ]$ is $c_n$.

In the evaluation of the average $\mathbb E\left[\,pf[ \mathcal A]\,\cdot \,pf[ \mathcal A^T] \,\right]$ the only non-zero terms are the $c_n$ terms that are product of entries symmetric 
with respect of the matrix diagonal.  $\Box$\\

{\bf Proof 2 (Grassmann integral).} We compute the ensemble average of the characteristic polynomial $p(z)= \mathbb E  [\det \left (z\, I_n - {\mathcal A} \right ) ]$,
via a representation of the determinant of a matrix as a Gaussian integral on anti commuting variables ${\bar \psi}_i$ , $\psi_i$ , $i=1,\dots , n$.  
 \begin{align*}
\det \left(z\,I_n-{\mathcal A}\right)=& \int \prod_{k=1}^n  \left(d{\bar \psi}_k d\psi_k\right) \,e^{\sum_{i,j} {\bar \psi}_i\left(z\,I-
 {\mathcal A} \right)_{i,j} \psi_j}=\\
=&\int \prod_{k=1}^n  \left(d{\bar \psi}_k d\psi_k\right) \, e^{z \sum_{r=1}^n {\bar \psi}_r\psi_r}\,e^{-\sum_{i<j} {\bar \psi}_i {\mathcal A}_{i,j} \psi_j+ {\bar \psi}_j {\mathcal A}_{j,i} \psi_i} = \\
=& \int \prod_{k=1}^n\left(d{\bar \psi}_k d\psi_k\right) \,\prod_{r=1}^n \left(1+z\,{\bar \psi}_r\psi_r \right) \times \\
&\times \prod_{i<j} \left(1-{\bar \psi}_i {\mathcal A}_{i,j} \psi_j - {\bar \psi}_j {\mathcal A}_{j,i} \psi_i-{\bar \psi}_i\psi_i {\bar \psi_j}\psi_j\,{\mathcal A}_{i,j}{\mathcal A}_{j,i}\right) 
\end{align*}
The ensemble average is taken, with $\theta_{i,j}=-\mathbb E [{\mathcal A}_{i,j} {\mathcal A}_{j,i}]\ge 0$:
\begin{align*}
p(z) =& \int \prod_{k=1}^n\left(d{\bar \psi}_k d\psi_k\right) \,\prod_{r=1}^n \left(1+z\,{\bar \psi}_r\psi_r \right) \,
 \prod_{i<j} \left(1+{\bar \psi}_i\psi_i {\bar \psi_j}\psi_j\, \theta_{i,j}\right) =\\
 =& z^n +  z^{n-2}\sum_{i<j} \theta_{i,j}+ z^{n-4} \sum^\prime_{i_1<j_1\, , \, i_2<j_2} \theta_{i_1,j_1}\theta_{i_2,j_2} + \\
  &  + z^{n-6} \sum^\prime_{i_1<j_1\, , i_2<j_2 \, , \, i_3<j_3} \theta_{i_1,j_1}\theta_{i_2,j_2}\theta_{i_3,j_3} +\ldots ;
\end{align*}
if $n$ is even, the sum terminates with 
\begin{align}
\mathbb E[\det \mathcal A]= \sum^\prime_{i_k<j_k } \theta_{i_1,j_1}\dots \theta_{i_{n/2} ,j_{n/2}  }. 
\end{align}
The primed sums are restricted to have all indices different and $i_1<i_2<\dots <i_k$.
$\Box$

\rem{PPPPPP
\begin{prop}
Let $x_k+iy_k$ be the $n$ eigenvalues of a real matrix $M = D+gA$, where $D$ is diagonal with elements in 
$[-a,-b]$ and $A=-A^T$. For large $g$ the values $x_k$ condense to $\bar d$, where:
$$ \bar d = \sum_{j=1}^n d_j p_j, \qquad p_j =\frac{\sum_k (A_{j,k})^2}{\sum_{kj} (A_{j,k})^2 }  $$ 
\begin{proof}
The traces of $M^2$ and of $M^3$ give: 
\begin{align}
&\sum_k x_k^2-y_k^2 = {\rm tr}(D^2)  - g^2 {\rm tr} (A^TA)\label{second}\\
&\sum_k x_k^3 -3x_ky_k^2 = {\rm tr}(D^3)  - 3g^2 {\rm tr} (A^T DA) \label{third}
\end{align}
Note that ${\rm tr}(A^TDA) = \sum_{i,j} d_j (A_{j,i})^2 = \bar d \;{\rm tr}(A^TA)< 0$. We may normalize $g$ as to put
${\rm tr}(A^TA)=1$.
While the values $x_k$ remain bounded in $[-a,-b]$ for all $g$, the equations show that $y_k^2$ and $x_k y_k^2$ on average grow quadratically in $g$:
$$  \sum_k y_k^2 \: \approx\:  g^2, \qquad  \sum_k x_k y_k^2 \;\approx \; g^2 \bar d $$
meaning that $x_k\to \bar d$.\\
By the central limit theorem, for large order $n$ of the matrices: $\bar d=\bar x =\frac{1}{n} {\rm tr}D$. 
\end{proof}
\end{prop}
PPPPP}

\section*{Acknowledgments}
 Amos Maritan suggested  G.M.C. to study this problem and Enrico Onofri 
 helped with simulations and relevant remarks.\\

\begin{thebibliography}{99}
%
\bibitem{seli}  T.~Seligman, J.~Verbaarschot and M.~Zirnbauer, J. Phys. A: Math.
Gen. {\bf 18} (1985) 2751.
%
\bibitem{cas} G.~Casati, L.~Molinari, and F.~Izrailev, \textsl{Scaling properties of band random matrices}, Phys. Rev. Lett. {\bf 64} (1990),  1851--1854. G.~Casati, F.~Izrailev and L.~Molinari, J. Phys. A: Math. Gen. {\bf 24} (1991) 4755.
%
 \bibitem{chir} G.~Casati, B.~V.~Chirikov, I.~Guarneri, F.~M.~Izrailev, \textsl{Band-random-matrix model for quantum localization in conservative systems}, Phys. Rev. {\bf E 48} (1993), R1613.
%
\bibitem{spe} T.~Spencer, \textsl{Random Band Matrices and Random Sparse Matrices}, chapter in \textsl{Handbook on Random Matrix Theory}, Editors: G. Akemann, J. Baik, and Ph. Di Francesco. Oxford University Press, 2011.
%
\bibitem{raj} K.~Rajan and L.~F.~Abbott, \textsl{Eigenvalue Spectra of Random Matrices 
for Neural Networks}, Phys. Rev. Lett. {\bf 97}, 188104 (2006).
%
\bibitem{amir} A.~Amir, N.~Hatano and D.~R.~Nelson, \textsl{Localization in non-Hermitian chains with excitatory/inhibitory connections}, arXiv:1512.05478.
%
\bibitem{all} 
S.~Allesina and Si Tang, 
\textsl{The stability -- complexity relationship at age 40: a random matrix
perspective}, Popul. Ecol. {\bf 57} (2015), 63--75.
%
\bibitem{may} R.~M.~May, \textsl{Will a large complex system be stable?}, Nature {\bf 238}, 413--414 (1972).
%
 \bibitem{cha} C.~Bordenave and D.~Chafai, {\it Around the circular law}, Probability Surveys Vol. 9 (2012) 1--89.
%
\bibitem{horn} R.~A.~Horn and C.~R.~Johnson, \textsl{Matrix Analysis}, chapter 5 and 6, 
Cambridge Univ. Press 1985.
%
\bibitem{ESA} ESA Report, D.~U.~Hooper et al., {\it Effects of biodiversity on ecosystem functioning: a consensus of current knowledge}, Ecological Monographs, 75(1), 2005, 3--35, by the Ecological Society of America.
%
\bibitem{rob} A.~Roberts, \textsl{Stability of a feasible random system} , 
Nature {\bf 251} (1974) 607--608.
%
\bibitem{imp} P.~Kirk, D.~M.~Y.~Rolando, A.~L.~MacLean and M.~P.~H.~Stumpf , \textsl{Conditional random matrix ensembles and the stability of dynamical systems}, New J. Phys. {\bf 17} (2015) 083025.
%
\bibitem{hay} D.~T.~Haydon, {\it Maximally stable model ecosystem can be highly connected}, Ecology {\bf 81} (9), (2000) 2631--2636.
%
\bibitem{McC} K.~S.~McCann,  \textsl{The diversity - stability debate}, Nature {\bf 405} (2000), 228--233.
%
\bibitem{for} M.~A.~Fortuna, D.~B.~Stouffer, J.~M.~Olesen, P.~Jordano, D.~Mouillot,  
B.~R.~Krasnov, R.~Poulin and J.~Bascompte, \textsl{Nestedness versus modularity in ecological networks: two sides of the same coin?}, J. of Animal Ecology {\bf 79} (2010), 811--817.
%
\bibitem{hersh}  D.~Hershkowitz,  \textsl{Recent directions in matrix stability}, 
Linear Algebra Appl. {\bf 171} (1992) 161-186.
%
\bibitem{Fyo2015}
Y.~V.~Fyodorov and B.~A.~Khoruzhenko, {\em A nonlinear analogue of May-Wigner instability transition}, arXiv:1509.05737 [cond-mat.dis-nn] (sept.2015).
%
\bibitem{meh} M.~L.~Mehta, \textsl{Matrix Theory, Selected Topics and Useful Results}, chapt. 11, Les Editions de Physique 1989.
%
\bibitem{somm} H.~J.~Sommers, A.~Crisanti, H.~Sompolinsky and Y.~Stein, 
\textsl{Spectrum of large random asymmetric matrices}, Phys. Rev. Lett. {\bf 60} (1988), 1895.
%
\bibitem{nau} A.~Naumov, \textsl{Elliptic law for real random matrices}, arXiv:1201.1639.
%
\bibitem{ngu}  Hoi~H.~Nguyen and S.~O'Rourke, \textsl{The elliptic law}, 
Int. Math. Res. Notices (2015) Vol. 2015, 7620--7689.  
%
\bibitem{wood} P. Matchett Wood, \textsl{Universality and the circular law for sparse random matrices}, Ann. Appl. Probab. {\bf 22}, No.3 (2012), 1266--1300.
%
\bibitem{rod} G.~J.~Rodgers and A.~J.~Bray, \textsl{Density of states of a sparse random matrix} ,
Phys. Rev. {\bf B 37}  (1988), 3557.  G.~J.~Rodgers and C.~De~Dominicis, 
\textsl{Density of states of sparse random matrices}, J. of Phys. A: Math. Gen {\bf 23}  (1990), 1567.
 %
\bibitem{fyo} Y.V. Fyodorov and A.D. Mirlin, \textsl{On the density of states of sparse random
matrices}, J. of Phys. A: Math. Gen. {\bf 24 } (1991), 2219.
 %
\bibitem{bauer} M.~Bauer and O.~Golinelli, \textsl{Random incidence matrices: moments of the spectral density}, J. Stat. Phys. {\bf 103} (2001), 301.
 %
\bibitem{seme} G.~Semerjian and L.~F.~Cugliandolo, \textsl{Sparse random matrices: the eigenvalue spectrum revisited}, J. of Phys. A: Math. Gen. {\bf 35} (2002), 4837.
%
\bibitem{kuhn} R.~K\"uhn, \textsl{Spectra of sparse random matrices}, 
J. Phys. A: Math. Theor. {\bf 41} (2008) 295002 (21pp).
%
\bibitem{tim} T.~Rogers, \textsl{New Results on the Spectral Density of Random Matrices}, 
Ph.D. thesis, Dept. of Math., King'\,s College London, July 2010
%
\bibitem{Metz}
I.~Neri and F.~L.~Metz, {\em Spectra of sparse non-Hermitian random matrices: an analytical solution}, Phys. Rev. Lett {\bf 109} (2012) 030602 (5pp).
%
\bibitem{bloch} J.~Bloch, F.~Bruckmann, N.~Meyer and S.~Schierenberg, 
\textsl{Level spacings for weakly asymmetric real random matrices and application to two-color QCD with chemical potential}, JHEP 08 (2012) 066. 
%
%
\end{thebibliography}
\end{document}